\documentclass{iaus}
\usepackage{graphicx}

\title[Insights from the Triangulum Spiral M33] 
{The Building of Galactic Disks: Insights from the Triangulum Spiral Galaxy
Messier 33}

\author[Block et al.]   
{David L. Block$^1$,
Iv\^anio Puerari$^2$,
Giovanni G. Fazio$^3$,
Alan Stockton$^4$,
Gabriela Canalizo$^5$,
Kenneth C. Freeman$^6$,
Thomas H. Jarrett$^7$,
Fran\c coise Combes$^8$
Robert Groess$^1$,
Guy Worthey$^9$,
Robert D. Gehrz$^{10}$,
Charles E. Woodward$^{10}$
and Elisha F. Polomski$^{10}$}

\affiliation{$^1$Anglo American Cosmic Dust Laboratory, School of Computational and Applied Mathematics, University
of Witwatersrand,
Johannesburg, 2050, South Africa \break email:
block@cam.wits.ac.za\\[\affilskip]
$^2$Instituto Nacional de Astrof\'\i sica, Optica y Electr\'onica,
Tonantzintla, 72840, M\'exico 
\break email: puerari@inaoep.mx\\[\affilskip]
$^3$Harvard-Smithsonian Center for Astrophysics, 60 Garden St.,
Cambridge,
MA 02138, USA \break email: gfazio@cfa.harvard.edu
\\[\affilskip]
$^4$Institute for Astronomy, University of Hawaii, 2680 Woodlawn Drive,
Honolulu, Hawaii, USA
\break email: stockton@ifa.hawaii.edu
\\[\affilskip]
$^5$Institute of Geophysics and Planetary Physics and Department of Physics,
University of California,
Riverside, CA 92521, USA
\\[\affilskip]
$^6$Mount Stromlo and Siding Spring
Observatories, Research School of Astronomy and Astrophysics, Australian
National University, Australia
\break email: kcf@mso.anu.edu.au\\[\affilskip]
$^7$Infrared Processing and Analysis Centre,
100-22, CALTECH, 770 South Wilson Ave, Pasadena, CA 91125, USA
\break email:
jarrett@ipac.caltech.edu
\\[\affilskip]
$^8$Observatoire de Paris, LERMA, 61 Av. de
l'Observatoire, F-75014, Paris, France
\break email:
francoise.combes@obspm.fr
\\[\affilskip]
$^9$Washington State University, 1245 Webster Hall, Pullman, WA 99163-2814,
USA
\break email: gworthey@wsu.edu
\\[\affilskip]
$^{10}$Department of Astronomy, University of Minnesota, 116 Church St. SE,
Minneapolis MN 55455, USA
\break
email:gehrz@astro.umn.edu}

\pubyear{2007}
\volume{235}  
\pagerange{1}
\date{?? and in revised form ??}
\jname{Galaxy Evolution Across the Hubble Time}
\editors{F. Combes \& J. Palous, eds.}
\begin{document}

\maketitle

\begin{abstract}
The Triangulum Spiral Galaxy Messier 33 offers unique insights into
the building of a galactic disk. We identify spectacular arcs
of
intermediate age (0.6 Gyr $-$ 2 Gyr) stars in the
low-metallicity outer disk.
The northern arc spans  $\sim$120 degrees in azimuth and up
to 5 arcmin in width. The arcs are located 2-3 disk scale lengths
from the galaxy centre (where 1 disk scale length is equivalent
to 0.1 degrees in the V-band) and lie precisely where there is a
warp in the HI profile of M33. Warps and infall are inextricably
linked (Binney, 1992). We present  spectroscopy of  
candidate stars in the outer northern arc, secured using the
Keck I telescope in Hawaii. The target stars have estimated
visual magnitudes as faint as V$\sim$ 25m. Absorption bands of
CN are seen in all spectra reported in this review talk,
confirming their carbon star status. Also presented are PAH emissivity
radial profiles generated from IRAC observations of M33 using the
Spitzer Space Telescope. A dramatic change of phase in the $m$=2
Fourier component is detected at the domain of the arcs.     
M33 serves as an excellent example how the disks of
spiral galaxies in our Universe are built: as dynamically open
systems, growing
from the inward, outward.


\end{abstract}

\firstsection 
\section{Introduction}

Spiral galaxies appear to be dynamically open systems, whose disks are
still forming at the current epoch and which continue to accrete mass.
Isolated spiral galaxies (non-accreting systems) cannot
reproduce the observed distribution of gravitational bar torques at
all (see Figure 1 in Block et. al. \cite{blocketal02}). In that study, it
was argued that disk masses should double every 10 billion years. The
most likely source for the origin of accreted gas could be the
reservoirs of gas observed outside nearly all spiral disks (Sancisi
\cite{sancisi83}). Keres et al. (\cite{keresetal05}) carefully discuss
accretion in terms of filaments in the cosmic web. An extremely
important point is that {\it accreting systems
in gas need not show any signs of accretion in stars}, such as the presence
of tidal tails,
stellar loops or close companions.
Disks of
galaxies appear to form from the inside out (Block et al. \cite
{blocketal02}).
Relative to the inner regions of spiral galaxies, the mean ages of
the outer regions are known to be somewhat younger and more
metal-poor (Bell and de Jong \cite{belldejong00}). We can therefore expect
the contribution
from intermediate-age (0.6 Gyr $-$ 2 Gyr) stars to be stronger in these outer
regions, and the near-IR surface brightness of the outer disk will
be preferentially enhanced by the presence of thermally-pulsing
asymptotic giant branch (TP--AGB) stars. Although the average number
of carbon stars per intermediate age cluster in the Magellanic clouds
is only about 2.5 (Persson et al. \cite{persson83}), they radiate such
an enormous amount of light in the near-infrared that they contribute
about 50 percent of the bolometric luminosity of the entire cluster.

There is no better place to start examining the accretion of gas in
the outer domains of spiral disks than in our Local Group, and we
focus our attention here on the Triangulum Spiral Galaxy M33. The
history of the Andromeda Galaxy M31 is discussed elsewhere (Block et
al. \cite{blocketal06}). If
accretion of gas is indeed taking place in M33, and if its disk is
growing from the inside outwards, then one obvious tell-tale signature would
be the presence of  a red intermediate age  
population of carbon stars in its
outer disk. 

In an earlier study (Block et al. \cite{blocketal04}),
we presented  near-infrared images of M33 from a
deep subsample of 2MASS, in which individual stars were unresolved.  
The deep 2MASS images revealed remarkable arcs of red stars in the
outer disk; the northern arc  subtends  $120^\circ$ in azimuth angle
and $\sim 5'$ in width (see Figure \ref{m33_2mass}).  The northern
arc is dominant although a very faint southern counterpart arc, forming
a partial ring, can also be seen. The arcs lie at a radius of 2-3 disk
\hbox{scale} lengths (in V, the disk scale length is 0.1 degree; Ferguson et
al.
\cite{fergusonetal06}).
Of course we are not claiming that carbon stars lie only in the two
outer arcs, but we do claim that a combination of increased number
density, perhaps aided by an increase in the IR integrated-light
contribution of carbon stars relative to the underlying stellar
population due to a metallicity gradient, makes them stand out in the
infrared. That there is an increase in number density at the appropriate
radius is confirmed by Rowe et al. (2005).

Fourier analysis of the
light distribution shows that the dominant $m=2$ peak
corresponds to the giant arcs.  The pitch angle of the dominant $m=2$
structure is $\sim$  58 degrees.  It is these outer arcs, not the
inner two-armed structure, which dominate the $m=2$ Fourier
spectra at the NIR.



\begin{figure*}[ht]
\vspace{12.0cm}
\caption{A partial ring of very red stars is seen in this JHKs image
of M33, secured from a special set of 2MASS observations
wherein the integration time was increased by a factor of six. The
ring is seen to full advantage with a simple ellipsoid model
subtracted.
The northern plume or arc spans up to $5'$ in width and is
located at a radius of 2 to 3 disk scale lengths.
An ellipsoidal swath is selected to pass through the arcs. North is up and
East to the left.} 
\label{m33_2mass}
\end{figure*}

\begin{figure*}[ht]
\vspace{9.0cm}
\caption{The positions of the seven stars listed in Table \ref{table_mag_cstars} are indicated
  by plus signs; the stars lie in the northern swath of carbon bearing
  stars identified photometrically in an earlier study (Block et
  al. \cite{blocketal04}). 
 Spectra of
  these stars (see Fig. 3)
were secured using the Keck I telescope in Hawaii. Also observed
  (but with its twin sister telescope, Keck II) are two stars 
   which lie 0.5 degrees away from the centre of M33, in an area
  identified by a white circle. These outlying stars are representative of a
  very red family identified in near-infrared imaging with the Hale
  5m reflector at Mount Palomar.}
\label{m33_plus_carbonstars}
\end{figure*}

\begin{figure*}[ht]
\vspace{12.0cm}
\includegraphics{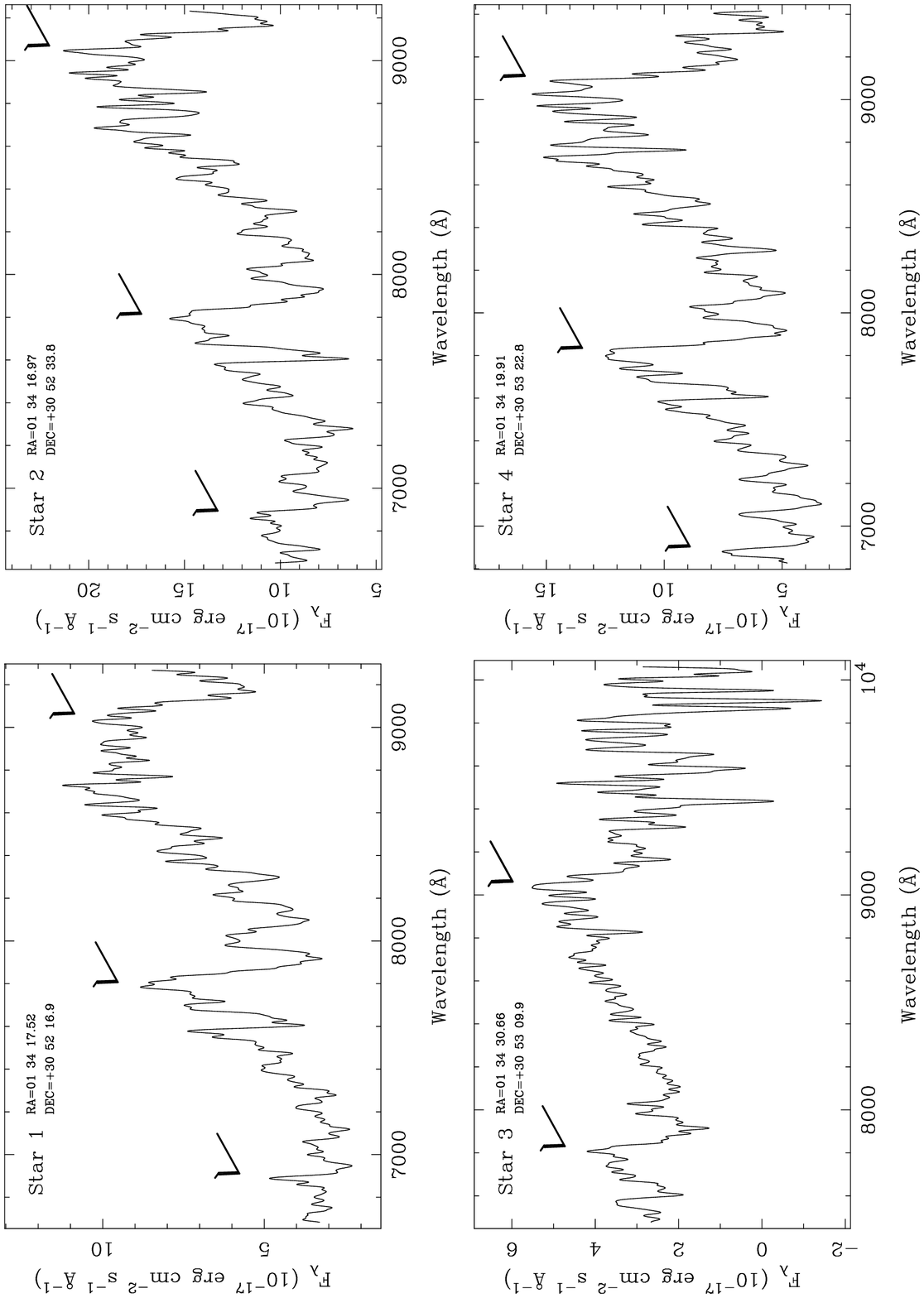}
\caption{Spectra of four stars in the northern swath seen in Figure \ref
{m33_2mass},
secured using the Keck
 I telescope.
The coordinates of each star are indicated in each
 panel. Carbon stars show a
plethora of molecular spectral features, including the C$_{2}$ Swan
bands and the CN
bands. The presence
of the hugely dominant CN bands shortward of 7000\AA\  and 8000\AA\
and longward of 9000\AA\ are indicated by tick marks 
and unambiguously reveal their C-star status. }
\label{panel_four}
\end{figure*}

\section{Observations}


Table \ref{table_mag_cstars} lists the magnitudes and colors of 7 candidate
targets in the outer northern
arc. In deriving the absolute
magnitudes in Table \ref{table_mag_cstars},
we assume a distance modulus to M33 of
24.64$^{m}$, corresponding to a linear distance of 840 kpc (Freedman
et al. \cite{freedmanetal91}). Noting that \hbox{E(B$-$V) = 0.3 $\pm$ 0.1
mag}
(Wilson \cite{wilson91}), and that the K-band extinction is approximately
one-tenth that in the optical, 
we
use a
dust extinction correction for the Ks apparent magnitudes of 0.09 mag.

\begin{table}
 \caption[]{Magnitudes and colors of our target stars in the northern red
arc of M33}
  \begin{center}
\begin{tabular}{cccccccc}
\hline
Star & J  &  H &     Ks   &   J$-$Ks  &  J$-$H   & H$-$Ks & M$_{\rm Ks}$ \\
\hline
1 & 17.96 & 16.76 & 15.91 &  2.05  & 1.20  & 0.85 & $-$8.81 \\
2 & 18.49 & 17.09 & 16.32 &  2.16  & 1.40  & 0.77 & $-$8.39 \\
3 & 19.61 & 18.04 & 16.66 &  2.96  & 1.57  & 1.39 & $-$8.06 \\
4 & 18.57 & 17.26 & 16.49 &  2.08  & 1.31  & 0.77 & $-$8.23 \\
5 & 18.82 & 17.35 & 16.54 &  2.28  & 1.47  & 0.81 & $-$8.17 \\
6 & 18.42 & 17.24 & 16.46 &  1.97  & 1.18  & 0.78 & $-$8.25 \\
7 & 18.37 & 16.91 & 15.82 &  2.55  & 1.46  & 1.10 & $-$8.89 \\
\hline
\end{tabular}
\end{center}
\label{table_mag_cstars}
\end{table}

Spectra of our carbon star candidates were obtained on 2004 Aug 17
UT with the Low Resolution Imaging Spectrograph (LRIS; Oke et al.
\cite{okeetal95}) attached to the Keck I telescope.

Observations of M33 were also made using all four bands of the Infrared
Array Camera (IRAC) on 2005 January 21 as part of a
Spitzer Guaranteed Time Observing Program (Program ID 5) conducted by
Spitzer Science Working Group member R. D. Gehrz.
The IRAC instrument (Fazio et al. \cite{fazioetal04}) is composed of four detectors
which operate at 3.6, 4.5, 5.8, and 8.0 $\mu m$.  All four detector arrays
are 256x256 pixels in size with mean pixel scales of 1.221, 1.213, 1.222,
and 1.220$'' \rm{pixel}^{-1}$ respectively.  





\begin{figure}
\vspace{11.0cm}
\includegraphics{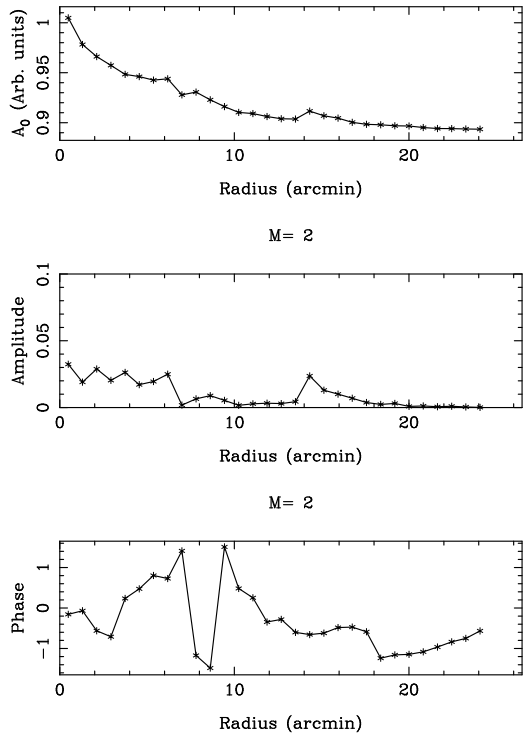}
\caption{The topmost panel shows a PAH emissitivity radial
  profile. In the middle panel is shown the amplitude of the Fourier
  m=2 mode as a function of radius. The local peak at $r=14'$
  coincides with the inner radial limit of the arcs. The phase of the
  m=2 component is
  presented in the lower panel. We see a dramatic change of
  phase in the PAH distribution at precisely 
 the same radius $r=18'$ where Corbelli and Schneider
  (\cite{corbellischneider97}) find a warp in HI. }
\label{m33_ampli_phase}
\end{figure}



\section{Discussion and Conclusions}

Carbon stars show a
plethora of molecular spectral features, including the C$_{2}$ Swan 
bands and the CN
bands. All stars in Table 1 as well as those imaged
within the white circle (Figure 2) are spectroscopically confirmed to be carbon stars.

 Figure 4 shows a PAH emissitivity radial
  profile in M33, generated from a scaled non-stellar 
IRAC 8$\mu m$ - $3.6\mu m$
  image.  We find a dramatic change of
  phase in the PAH distribution at precisely 
 the same radius where Corbelli and Schneider
  (\cite{corbellischneider97}) find a warp in HI. Galactic warps and
  infall of gas are inextricably linked, as reviewed by Binney (1992).

The Keck spectra offers unique insight into how the disks of one of our
closest spiral
galaxies, M33,  has evolved with time:
it continues to grow by accretion of low-metallicity gas from the
inside, outwards.  
The implications of the presence
of carbon
stars  in the outer disk of M33
immediately beckons the question of a possible ubiquity of such an
intermediate age population in the outer domains of other spiral
disks. 
At hand is the potential to
exploit our technique of identifying thermally-pulsing
 asymptotic giant branch carbon stars in the outer domains of 
more distant spirals on the basis
 of their near-infrared colours. 

\acknowledgements{
The Anglo American Cosmic Dust Laboratory is funded by the 
Anglo American Chairman's Fund. 
IP acknowledges support from the Mexican foundation CONACyT
under project 35947--E. 
The paper is based in part on data obtained at the W. M. Keck
Observatory. The Spitzer Space Telescope is operated by the JPL, Caltech 
under a contract with
NASA. RDG and CEW were supported by NASA through an award
issued by JPL/Caltech.}

\end{document}